%% file: main.tex
\documentclass[conference]{IEEEtran}
\newlength{\flexwidth}
\usepackage[top=0.75in, left=0.625in, right=0.625in, bottom=1in]{geometry}
\usepackage{amsmath,amssymb,amsthm,fixmath}
\usepackage{mathtools}
\usepackage{accents}
\usepackage{xcolor}
\usepackage{siunitx}
\usepackage{cuted}
\usepackage{multirow,booktabs}
\usepackage{subfigure}
\usepackage[inline]{enumitem}
\usepackage{optidef}
\usepackage{graphicx}
\usepackage{lipsum}
\usepackage{amssymb}
\usepackage{diagbox}
\usepackage{adjustbox}
\pdfoutput=1
\usepackage[linesnumbered,titlenumbered,ruled,vlined,resetcount,algosection]{algorithm2e}

\usepackage[hidelinks]{hyperref}

\usepackage{epstopdf}
\usepackage[textsize=tiny,colorinlistoftodos]{todonotes}

\usepackage[shortcuts,acronym,automake]{glossaries}
\input{glossary}

\DeclareSIUnit{\belmilliwatt}{Bm}
\DeclareSIUnit{\dBm}{\deci\belmilliwatt}

\usepackage{tcolorbox}
\tcbuselibrary{many}

\usepackage[free-standing-units=true]{siunitx}

\begin{document}

	\title{A Reliable and Resilient Framework for Multi-UAV Mutual Localization}
	
	\author{
		\IEEEauthorblockN{Zexin~Fang\IEEEauthorrefmark{1},~Bin~Han\IEEEauthorrefmark{1}~and~Hans~D.~Schotten\IEEEauthorrefmark{1}\IEEEauthorrefmark{2}}
		\IEEEauthorblockA{
		\IEEEauthorrefmark{1}Rheinland-Pf\"alzische Technische Universit\"at Kaiserslautern-Landau  (RPTU), Kaiserslautern, Germany\\
		\IEEEauthorrefmark{2}German Research Center of Artificial Intelligence (DFKI), Kaiserslautern, Germany
		}
	}
	
	\bstctlcite{IEEEexample:BSTcontrol}
	
	\maketitle
	
	\begin{abstract}
		This paper presents a robust and secure framework for achieving accurate and reliable mutual localization in multiple \gls{uav} systems. Challenges of accurate localization and security threats are addressed and corresponding solutions are brought forth and accessed in our paper with numerical simulations.
        The proposed solution incorporates two key components: the \gls{magd} and \gls{tad}. The \gls{magd} adapts the gradient descent algorithm to handle the configuration changes in the mutual localization system, ensuring accurate localization in dynamic scenarios. The \gls{tad} cooperates with \gls{rp} scheme to detect and mitigate potential attacks by identifying \gls{uav}s with malicious data, enhancing the security and resilience of the mutual localization.     
	\end{abstract}
	
	
	\IEEEpeerreviewmaketitle
	
	\section{Introduction}\label{sec:introduction}
    Multi-\gls{uav} systems hold significant promise for revolutionizing various domains, particularly the future Sixth Generation (6G). For instance, multi-\gls{uav} systems have the potential to provide reliable communication links in challenging environments, support connectivity in remote or disaster-stricken areas, and overcome limitations of ground infrastructure \cite{FengSCE}.

    In \gls{uav} applications, \gls{gps} modules may face challenges in providing precise position information in urban areas, tunnels, or environments with obstacles. Alternative options like radio trilateration can be used but have limited coverage and require calibration and installation cost \cite{5Gposition}\cite{Radiotri}. A mutual position system utilizing anchor \gls{uav}s with accurate \gls{gps} positions can provide precise estimates for target \gls{uav}s with poor \gls{gps} reception. Distance estimation methods such as \gls{tof} or \gls{rssi} ranging can be used in this system.


	Existing localization research primarily addresses terrestrial scenarios in static networked sensors, often lacking altitude considerations. However, the uneven distribution of anchor \gls{uav}s in three dimensions adds complexity to localization algorithms. Additionally, in multi-\gls{uav} mutual localization, the mobility of \gls{uav}s introduces variations in reliable position and distance information. Therefore, comprehensive research and novel methodologies are necessary to tackle these challenges \cite{movinglocal}.
    On top of that, security is also a critical concern in multi-\gls{uav} mutual localization. While extensive research has been done on security in static sensor networks \cite{Ragmalicous}\cite{Jhagame}\cite{Tomic2022DetectingDA}, its validation in dynamic scenarios is lacking. The changing topology of target \gls{uav}s and malicious \gls{uav}s can significantly impact the performance of attack and defense schemes. Further investigation is needed to validate the performance and robustness of these schemes for dynamic multi-\gls{uav} mutual localization scenarios.
    
    This work demonstrates a robust and secure framework that ensures accurate and reliable mutual localization while addressing potential security threats. The subsequent sections of the paper are structured as follows. In Sec.\ref{Errmodel} we investigate the error model of our scenario. In Sec.\ref{MAGD} we introduce our proposed methodologies, meanwhile, evaluate the overall efficiency and accuracy of the mutual localization system in the specific scenario under consideration. Then in Sec\ref{attack} we outline the potential attack scheme and proposed defend scheme (\gls{tad}), and validate our proposed scheme with numerical simulations presented in Sec.\ref{SIMUTDAD}. In Sec.\ref{conclu}, we conclude our paper by summarizing the key findings.

	\section{Error model setup}\label{Errmodel}
    \subsection{Distance estimation error}\label{disERR}
    Considering the limited range and sensitivity to environmental conditions of \gls{tof} ranging, a better approach that suits our scenario is \gls{rssi} ranging. This method relies on a path loss model to establish a relationship between the \gls{rssi} value and the distance of a radio link from an anchor \gls{uav}. Such a model can be described as:
    \begin{equation}\label{eq:pathloss}
		P_\mathrm{r}\textit{(d)}= P_\mathrm{r}(\textit{d}_0) - 10n_\mathrm{p} * \mathrm{log}(\frac{\textit{d}}{\textit{d}_0}),
	\end{equation}
    where $P_\mathrm{r}\textit{(d)}$ indicates the \gls{rssi} value at the distance $\textit{d}$ from the anchor \gls{uav}; $\textit{d}_0$ is the predefined reference distance, meanwhile $P_\mathrm{r}(\textit{d})$ is the \gls{rssi} measurement value at $\textit{d}_0$; $n_p$ denotes the path loss factor of the radio link.
    
    Regarding the fact that a radio channel suffers from different forms of fading, therefore \gls{rssi} measurement error $P_\mathrm{r}^{\Delta}(\textit{d})$ is inevitable and will result in a distance estimation error $\mathcal{E}$. We investigated \gls{rssi} measurement results between two Zigbee nodes \cite{shortrangenovel} \cite{shortrangeoutdoor} and Sigfox nodes \cite{longseparation}. $P_\mathrm{r}^{\Delta}(\textit{d})$ is zero-mean Gaussian distributed with a standard deviation $\sigma_\mathrm{r}{(\textit{d})}$. However, in the case of small distances, $\sigma_\mathrm{r}{(\textit{d})}$ shows a weak relation to the distance, which indicates fading dominates over path loss in this range. We consider an outdoor urban environment with NLOS (Non-Line of sight) radio links, which is quite common in our use case, thus $P_\mathrm{r}^{\Delta}(\textit{d})$ is likely to be fluctuating in a same manner. To simplify our analysis, the standard deviation $\sigma_\mathrm{r}{\mathrm{(\textit{d},t)}}$ is considered to be randomly and uniformly distributed, meanwhile subjected to the square of the distance to approximate measurement results, as described follows:
    \begin{equation}\label{eq:errRSSI}
        P_\mathrm{r}^{\Delta }{\mathrm{(\textit{d},t)}}\sim\mathcal{N}\left(0,\sigma_\mathrm{r}{\mathrm{(\textit{d},t)}}^2\right)
    \end{equation} 
    \begin{equation}\label{eq:deviationRSSI}
        \sigma_\mathrm{r}{\mathrm{(\textit{d},t)}}\sim\gamma_d*\mathcal{U}\left(\sigma_{\mathrm{min}}^r,\sigma_{\mathrm{max}}^r\right)
	\end{equation}
    \begin{equation}\label{eq:scale}
        \gamma_d =  S * \textit{d} ^2 +1
	\end{equation}
    where $\gamma_d$ is the modification factor, which modifies the relation between distance and $\sigma_\mathrm{r}{\mathrm{(\textit{d},t)}}$; $S$ is the scaling factor, which is to scale how strong is $\gamma_d$ subjected to $\textit{d}$.
    
\begin{figure}[!htbp]
		\centering
		\includegraphics[width=0.90\linewidth]{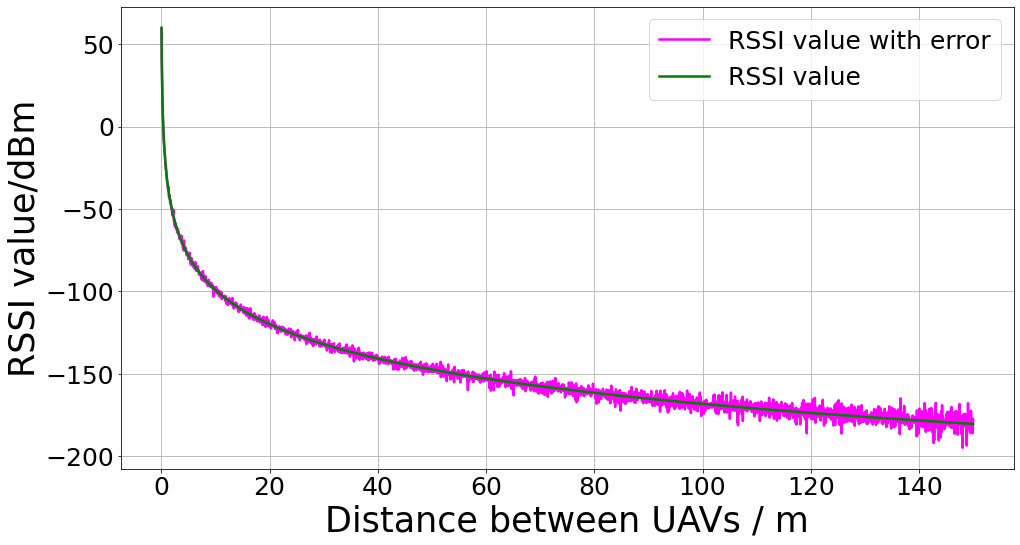}
		\caption{Simulated \gls{rssi} value over distance}
		\label{fig:RSSImean}
	\end{figure}

    Assume a pass loss model with pass loss factor $n_\mathrm{p} = 3$, reference distance $\textit{d}_0 = 1$ and \gls{rssi} measurement value $P_\mathrm{r}(\textit{d}_0) = -30 \si{\dBm}$, while $\sigma_{\mathrm{min}}^r = 0.5$, $\sigma_{\mathrm{max}}^r = 2$ and $S = 0.0001$. The simulation of such a model is shown in Fig.~\ref{fig:RSSImean}.
 

    
\begin{figure}[!htbp]
		\centering
		\includegraphics[width=0.90\linewidth]{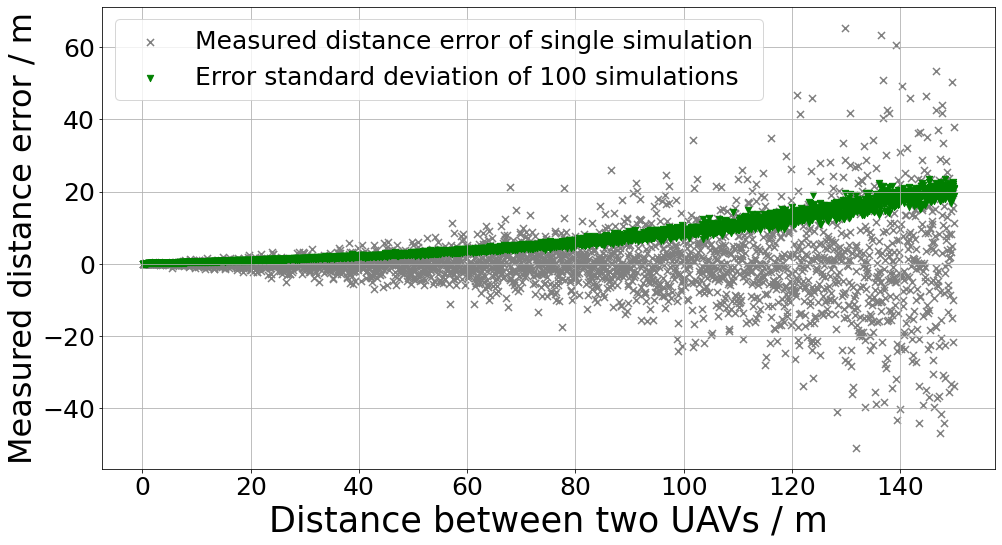}
		\caption{Estimation error and standard deviation over distance between 2 \gls{uav}s}
		\label{fig:RSSI100}
	\end{figure}

    By applying error to $P_\mathrm{r}(\textit{d})$ and find the corresponding distance of errored \gls{rssi} value, we are able to estimate the distance error $\mathcal{E}(\textit{d})$ over the distance, simulation results are presented in Fig.~\ref{fig:RSSI100}
 
    The simulation results shows that distance estimation error $\mathcal{E}(\textit{d})$ is zero mean Gaussian distributed with $\mathcal{E}(\textit{d})\sim(0,\sigma_\textit{d}^2(\textit{d}))$, while $\sigma_\textit{d}(\textit{d})$ fluctuates but increases over the distance. However, based on the simulation results, distance estimation over $150\meter$ can be extremely unreliable. Additionally, the communication cost can be drastically increased as more anchor \gls{uav}s are exposed to the target \gls{uav}.
    

   \subsection{Position measurement error}
   For a set of anchor \gls{uav}s  $\mathcal{U} = \{u_0,u_1,\dots u_\mathrm{n}\}$, each \gls{uav} has an error of its position, typically caused by various factors. With the assumption that the \gls{gps} module of \gls{uav} mitigated many of these error factors, we limited the \gls{gps} error to the Gussian error caused by the measurement. The actual position $\mathrm{p}_n(t)$ and its pseudo position $\mathrm{p^{\circ}}_n(t)$ at time step $t$ of the $n_\mathrm{th}$ \gls{uav} can be described : 
   \begin{equation}
        \mathrm{p}_n(t) = [x_n(t),y_n(t),z_n(t)]
    \end{equation}
    \begin{equation}
        \mathrm{p^{\circ}}_n(t) = \mathrm{p}_n(t) + [x_n^{\Delta}(t), y_n^{\Delta}(t),z_n^{\Delta}(t)]
    \end{equation}    
   \begin{equation}
		[x_n^{\Delta}(t), y_n^{\Delta}(t),z_n^{\Delta}(t)]\sim\mathcal{N}^2\left(0,\sigma_{\mathrm{p},n}^2/3\right)
   \end{equation}
   \begin{equation}\label{eq:lse}
        \sigma_{\mathrm{p},n}\sim\mathcal{U}\left(\sigma_{\mathrm{min}}^p,\sigma_{\mathrm{max}}^p\right)   
   \end{equation}
   where $[x_n^{\Delta}(t), y_n^{\Delta}(t),z_n^{\Delta}(t)]$ is position measurement error, which can be described by a zero mean Gaussian distribution with standard deviation $\sigma_{\mathrm{p},n}$.   $\sigma_{\mathrm{p},n}$ is considered to be random uniform distributed. Within a valid coverage of mutual localization, a target \gls{uav} $u_k\in\mathcal{U}$ measures the distance from anchor \gls{uav}s. The real distance $\textit{d}_{k,n}$ and measured distance $\textit{d}^{\circ}_{k,n}$ can be described as follows :
   \begin{equation}
	{d}_{k,n} = \left\Vert\mathrm{p}_k(t) - \mathrm{p}_n(t)\right\Vert
   \end{equation}
   \begin{equation}
		d^{\circ}_{k,n}= {d}_{k,n} + {\mathcal{E}({d}_{k,n},t)}
   \end{equation}
   \section{Adaptive and robust mutual localization}{\label{MAGD}}

   \subsection{Introduction to different localization techniques}{\label{introlocalt}}
   Thorough studies have been conducted focusing on distance-based localization techniques. We focus on 3 different localization techniques: \gls{ls}  based localization, \gls{ln-1} based localization, and \gls{gd} based localization, to investigate their localization performance in the presence of uneven spatial distribution of anchor \gls{uav}s. These techniques have been widely recognized for their robustness and efficiency in sensor network scenarios \cite{Ragmalicous}\cite{GVW2012efficient}.
   
   Given the measured distance ${d}^{\circ}_{k,n}$ of anchor \gls{uav}s and their positions $\mathrm{p^{\circ}}_n$, the position of $u_k$ can be estimated by minimizing the error between measured distance and calculated distance, as described follows:
   \begin{equation}\label{eq:optimize}
   \begin{split}
         {\mathrm{p}}_k &= \arg\min\limits_{[x,y,z]}\sum_{n=0}^{N}\left\vert\Vert\mathrm{p^{\circ}}_n-{\mathrm{p}}_k\Vert - {d}^{\circ}_{k,n}\right\vert.\\
    \end{split}     
   \end{equation} 
   This optimization problem can be directly solved by \gls{ls} technique with,
   \begin{equation}
   [x_k,y_k,z_k,\Vert\mathrm{p}_k\Vert^2]^\mathrm{T} = \mathrm{(A^TA)^{-1}Ab}   
   \end{equation} 
   where $\mathrm{A}$ and $\mathrm{b}$ are matrices containing anchor position and measured distances information,
   \[
   A = \begin{Bmatrix}
   -2x_0 \kern-6pt & -2y_0 \kern-6pt & -2z_0 \kern-6pt & 1 \\
   \vdots \kern-6pt & \vdots \kern-6pt & \vdots \kern-6pt & 1 \\
   -2x_n \kern-6pt & -2y_n \kern-6pt & -2z_n \kern-6pt & 1 \\
   \end{Bmatrix}
   \mathrm{b} = \begin{Bmatrix}
   {{d}^{\circ}}^2_{k,0} - \Vert\mathrm{p}^{\circ}_0\Vert^2 \\
   \vdots   \\
   {{d}^{\circ}}^2_{k,n} - \Vert\mathrm{p}^{\circ}_n\Vert^2 \\
   \end{Bmatrix}
   \]
   Moreover, such a localization problem can be formulated as a plane fitting problem, where the objective is to find a 4D plane $W = f(x, y,z)$ that fits the measurements $\mathrm{A}$ and $\mathrm{b}$. The coefficients of the plane can be represented as $u = [x_k,y_k,z_k,\Vert\mathrm{p}_k\Vert^2]^{\mathrm{T}}$. The optimization can be performed by minimizing the $l$1 norm-based distance metric \cite{Ragmalicous}. 
   \begin{equation}
   \begin{split}
      &\min\limits_{u,w}\Vert w\Vert_1\\
      &\text{subject to } \mathrm{A}u-w = \mathrm{b}
   \end{split}
   \end{equation}
   And $u$ can be solved iteratively by using \gls{admm} steps, 
   \begin{equation}
   \begin{split}
      u^i&= \mathrm{GA^T}(b+u^{i-1}-\frac{\lambda^{i-1}}{\rho})\\
      w_k&= S_{\frac{1}{\rho}}(\mathrm{A}u^{i-1}-b+\frac{\lambda^{i-1}}{\rho})\\
      \lambda^{i}&= \lambda^{i-1} + \rho(\mathrm{A}u^{i}-w^{i}-b)
   \end{split}
   \end{equation}
   where $\mathrm{G=(A^TA)}^{-1}$, $\lambda$ is the Lagrange multiplier and $\rho$ is the penalty parameter for violating the linear constraint. Meanwhile,  $S_{\frac{1}{\rho}}$ is the soft threshold function in $l1$ norm, defined as $S_{\frac{1}{\rho}} = \mathrm{sign}(x)* \max({\vert x\vert-\frac{1}{\rho}}, 0)$.

   Eq.~\ref{eq:optimize} can be also reformulated as $\mathrm{p}_k =\arg\min\limits_{[x,y,z]}f(x,y,z)$. By applying gradient descent to cost function $f(x,y,z)$, we are able to estimate the position $\hat{\mathrm{p}}_k$ of $u_k$ iteratively. At the $i_\mathrm{th}$ iteration, the negative gradient $g^{i}$ and position $\hat{\mathrm{p}}_k$ can be calculated,
   \begin{equation}
		g^{i} = - \nabla_{(x,y,z)} (f(x,y,z))\vert_{(x=\hat{x}_k^{i-1},y=\hat{y}_k^{i-1},z=\hat{z}_k^{i-1})}
   \end{equation}
   \begin{equation}
		\hat{\mathrm{p}}^i_k = \hat{\mathrm{p}}^{i-1}_k  + \alpha^{i} \times \frac{g^{i}}{||g^{i}||}
   \end{equation}
   where $\hat{\mathrm{p}}^{i-1}_k$ is the estimated position at the $(i-1)_{th}$ iteration. $\alpha^{i}$ is the step size at the $i_{th}$ iteration. $\alpha^{i}$ can be adjusted by discount factor $\beta$ to prevent over-descending. 

   The computational complexity of the three above-introduced techniques is summarized in Tab.~\ref{tab:compucomplex} \cite{Ragmalicous}. 
   \begin{table}[ht]\label{tab:compucomplex}
     \centering
     \caption{Computational complexity}
     \begin{tabular}{cccc}
     \toprule
      \gls{ls} & \gls{ln-1} & \gls{gd} \\
     \midrule
     $\mathcal{O}(N^2)$ & $\mathcal{O}(\max(K_{\mathrm{ADMM}}N,N^2))$ & $\mathcal{O}(K_{\mathrm{GD}}N)$\\
    
      \bottomrule
      \end{tabular}
    \end{table}
   \subsection{Performance evaluation under uneven spatial distribution}
\begin{figure}[!htbp]
		\centering
		\includegraphics[width=0.90\linewidth]{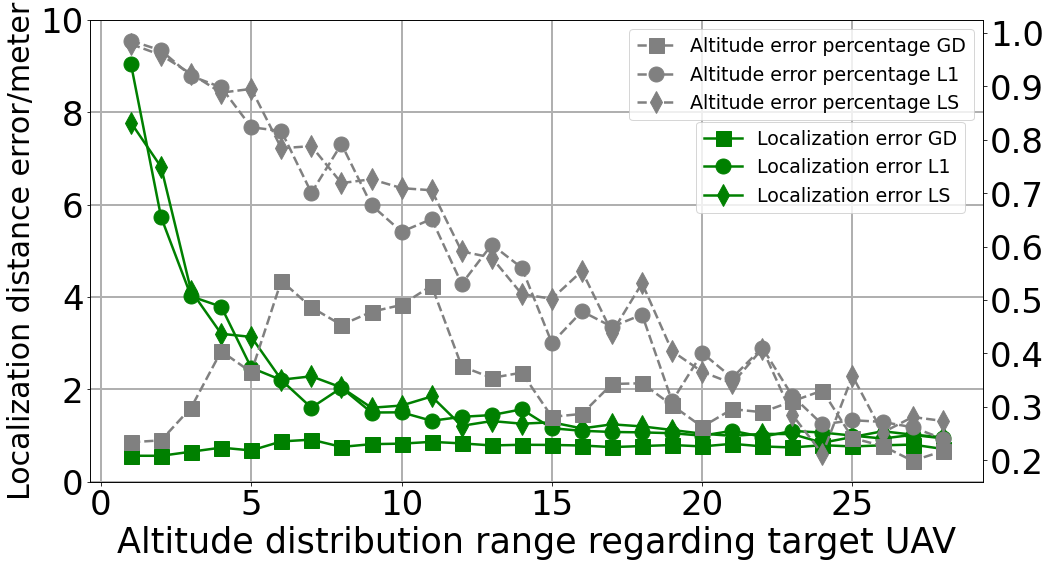}
		\caption{Localization error over spatial distribution}
		\label{fig:errorsp}
	\end{figure}

   For \gls{uav} applications, energy consumption is a critical concern. To compare the \gls{ls}, \gls{ln-1}, and \gls{gd} localization techniques under energy-limited conditions, we set $K_\mathrm{ADMM}=N$ and $K_{\mathrm{GD}} =N$ to ensure equal computational complexity for all three approaches. In the simulation, the target \gls{uav} $u_k$ is surrounded by anchor \gls{uav}s distributed within a cubic area. 
   The shape of this cubic area varies, ranging from $[x_k\pm35.35\metre,y_k\pm35.35\metre,z_k\pm1\metre]$ to $[x_k\pm29.29\metre,y_k\pm29.29\metre,z_k\pm28\metre]$ to keep a maximum $d_{k,n} = 50\metre$. The \gls{rssi} error model configured as shown in Fig.~\ref{fig:RSSImean}, and simulation is set up as follows: $N = 30$, $[\sigma_{\mathrm{min}}^p,\sigma_{\mathrm{max}}^p] = [0.1,3]$ and $\rho = 0.1$, $[\alpha^0,\beta] =[1,0.5]$. Each estimation was repeated 50 times to exclude randomness. The results are presented in Fig.~\ref{fig:errorsp}.
 
   The simulation results show that the error of \gls{ls} based and \gls{ln-1} based localization decreases as the distribution of anchor \gls{uav}s becomes more even in terms of latitude, longitude, and altitude. However, the error in \gls{gd} based localization is not significantly affected by the distribution of anchor \gls{uav}s. Specifically, when anchor \gls{uav}s are densely distributed in one dimension compared to the other two dimensions, \gls{ls} and \gls{ln-1} localization techniques can become highly unreliable in the densely distributed dimension. In contrast, \gls{gd} based localization doesn't show a strong correlation with the distribution of anchor \gls{uav}s in this regard. In summary, \gls{gd} based localization is better suited for scenarios involving multiple \gls{uav}s, where the distribution of \gls{uav}s may be uneven in all three dimensions. 
   \subsection{Weighted localization and error conversion}
\begin{figure}[!htbp]
		\centering
		\includegraphics[width=0.90\linewidth]{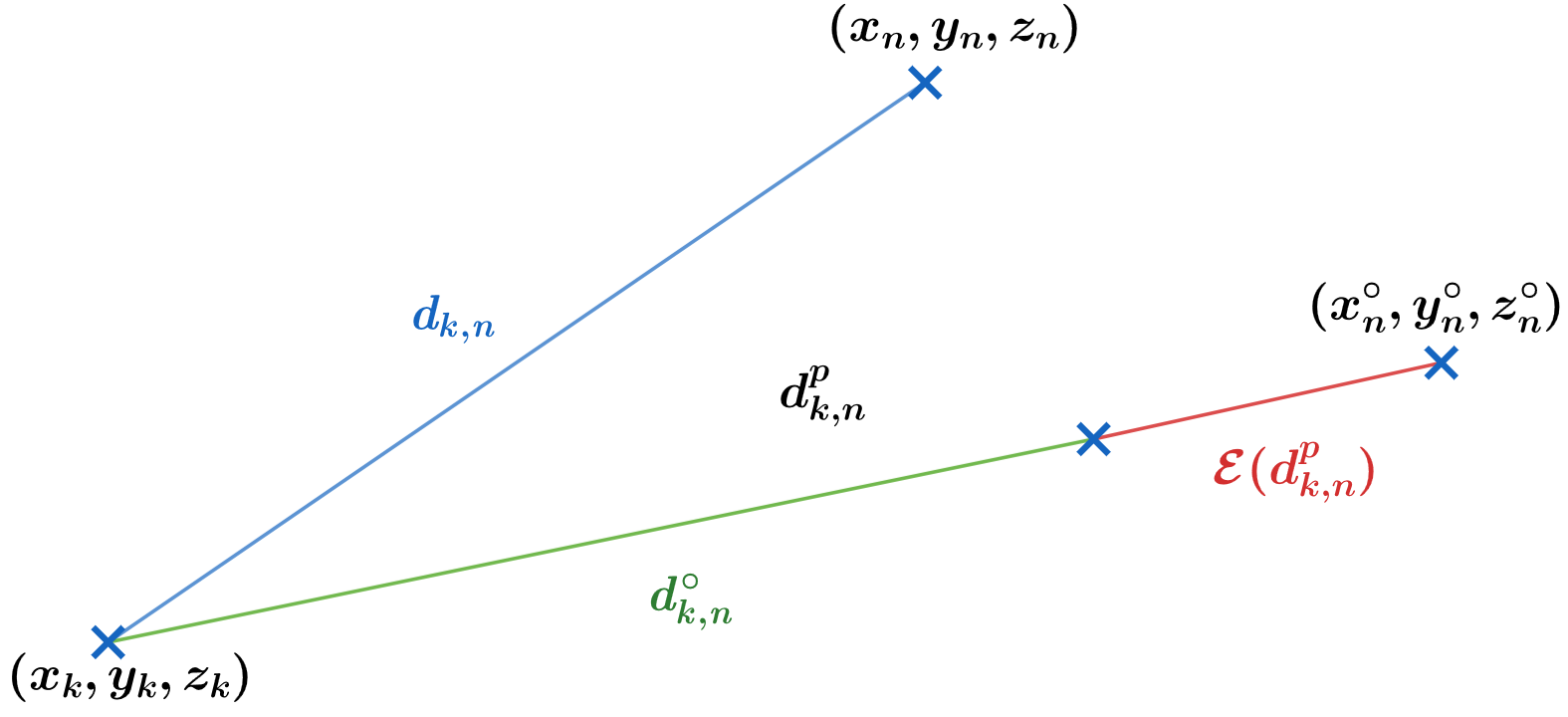}
		\caption{Position and distance estimation error conversion}
		\label{fig:ERRcon2}
	\end{figure}

   To speed up the convergence of gradient descent and ensure the stability of localization, a weighted localization approach can be proposed, as described below:  
    \begin{equation}
    \begin{split}
         \hat{\mathrm{p}}_k &= \arg\min\limits_{[x,y]}\sum_{n=0}^{N}\left\vert\Vert\mathrm{p}^{\circ}_n-\mathrm{p}\Vert - {d}^{\circ}_{k,n}\right\vert*w_n^e
    \end{split}     
    \end{equation}
    where $w_n^e$ is the error weight of anchor \gls{uav}s, which can be determined by the distance estimate error $\mathcal{E}$ and position measurement error power $\sigma_{\mathrm{p},n}$.
    To calculate the error weight, we can convert the position measurement error to the distance estimate error, as illustrated in Fig.~\ref{fig:ERRcon2}.
 
    For an anchor \gls{uav} $u_n$ with actual position $[x_n, y_n, z_n]$ and measured position $[x^{\circ}_n, y^{\circ}_n, z^{\circ}_n]$, $d_{k,n}$ is the actual distance and $d^{\circ}_{k,n}$ is the measured distance. $d^p_{k,n}$ is the erred distance solely caused by position error. $\mathrm{\mathcal{E}}(d^{p}_{k,n})$ is the converted distance estimation error jointly caused by position measurement and distance estimate. $\mathrm{\mathcal{E}}(d^{p}_{k,n})$ can be approximated by a non-zero mean Gaussian distribution, as introduced in \cite{han2023secure}.
    Such converted distance estimate error can described as $ \mathrm{\mathcal{E}}_{cd}\sim\mathcal{N}(\mu_{\mathrm{cd}},\sigma_{\mathrm{cd}})$. Nevertheless, $\sigma_{\mathrm{d}}(d_{k,n})$ can't be directly retrieved with a single measurement of $d^{\circ}_{k,n}$. With the approximation $\sigma_{\mathrm{d}}(d_{k,n}) \gets \sigma_{\mathrm{d}}(d^{\circ}_{k,n})$, $d^{\circ}_{k,n} \gets d_{k,n} + \mu_{cd}$ and known $\sigma_{\mathrm{p},n}$, $[\mu_{\mathrm{cd}},\sigma_{\mathrm{cd}}]$ can be obtained through least squares estimation. Error weight can be calculated as $w_n^e = \frac{\sum\limits_{u_n\in\mathcal{U}}\sigma_{\mathrm{c},n}}{n*\sigma_{\mathrm{c},n}}$.
    
    \subsection{Mobility adaptive gradient descent Algorithm}
    An appropriate initial step size $\alpha^0$ is crucial for gradient descent-based estimation. Conventional approaches with fixed initial step sizes may not perform optimally in our scenario. To tackle this issue, we can use adaptive step sizes that dynamically adjust at each stage based on the changing speed of the target \gls{uav} $u_k$ and the availability of anchor \gls{uav}s.

    \begin{algorithm}[!htbp]
    \caption{Mobility adaptive gradient descend algorithm}
    \label{alg:MAGD}
    \scriptsize
    \DontPrintSemicolon
    Input: $\hat{\mathrm{p}}(0) = \hat{\mathrm{p}}_{init}$; discount factor $\beta_1$ and $\beta_2$; step size threshold $\epsilon^{\mathrm{max}}_{\mathrm{t_0}}$,  $\epsilon^{\mathrm{min}}_{\mathrm{t_0}}$; maximum iteration, convergence threshold and momentum: $K, \theta, m $; Simulation time T

    Output: $\hat{\mathrm{p}}(t)$
    
    
    \SetKwProg{Fn}{Function}{ is}{end}
    \Fn{\textsc{MAGD}$(t = 1:T)$} {
        \If {$t = 1$}{$\hat{\alpha}\gets \max(\frac{\epsilon^{\mathrm{max}}_{\mathrm{t_0}}}{n},\epsilon^{\mathrm{min}}_{\mathrm{t_0}})$
        }
        
        update: $\hat{\mathrm{p}}\gets \hat{\mathrm{p}}(t-1); \bar{D}_0\gets +\infty$ 
        
        \For {$i = 1:I$ }
        {    \For {$n = 1:N$}{get $\mathrm{p}_n^{\circ}, \sigma_{\mathrm{p},n}, d^{\circ}_{n}$ from $u_n$
        
                   get $\mu_{\mathrm{c},n},\sigma_{\mathrm{c},n}$\tcp*{Error conversion}
                   
                   $w_n^e \gets \frac{\sum\limits_{u_n\in\mathcal{U}}\sigma_{\mathrm{c},n}}{n*\sigma_{\mathrm{c},n}}$  \tcp*{Calculate error weight}
                   
                   $\hat{d}_n = \Vert \hat{\mathrm{p}} - \mathrm{p}_n^{\circ}\Vert$ }
                   
             $G^i \gets \sum\limits_{u_n\in\mathcal{U}}\frac{(\hat{\mathrm{p}}-\mathrm{p}^{\circ}_n)*w_n^e }{\hat{d}_n}*(\hat{d}_n - d^{\circ}_n+\mu_{c,n})$ \tcp*{Gradient}
             
             update: $\hat{\mathrm{p}} \gets \hat{\mathrm{p}} + m*\hat{\mathrm{p}} + \frac{\hat{\alpha}}{n}* \frac{G^i}{\Vert G^i \Vert} $; $\hat{d}_n $
             $\bar{D}_i \gets \frac{1}{n} *\sum\limits_{u_n\in\mathcal{U}}(\hat{d}_n - d^{\circ}_n+\mu_{c,n})*w_n^e$ 
             
             \uIf{$\bar{D}_i > \bar{D}_{i-1}$}{
                $\hat{\alpha}\gets \hat{\alpha}*\beta_1$ \tcp*{reduce step size}
             }\ElseIf{$\frac{\bar{D}_{i-1} - \bar{D}_i}{\bar{D}_i} <= \theta$}{break}
        }
        update: $\hat{\mathrm{p}}(t)\gets \hat{\mathrm{p}}; \bar{D}(t)\gets \bar{D}_i$; 
                
        $V(t) \gets \Vert\hat{\mathrm{p}}(t) - \hat{\mathrm{p}}(t-1)\Vert$\tcp*{Estimate speed}

        $\bar{V} \gets \frac{1}{t}\sum\limits_{t=1}^{t}V(t)$\tcp*{Estimated average speed}

        $\Bar{\Bar{D}} \gets \frac{1}{t}\sum\limits_{t=1}^{t}D(t)$ \tcp*{Average  distance difference}

        
        \uIf {$t<\Phi$}{$\phi=t$}

        \ElseIf{$t>=\Phi$}{$\phi=\Phi$}
        
        $\rho_D \gets \frac{\sum\limits_{t=t-\phi}^{t}\bar{D}(t)}{(t=t-\phi)*\bar{D}}$ \tcp*{Apply smooth window}
        
        $\rho_V \gets \frac{\sum\limits_{t=t-\phi}^{t}V(t)}{(t=t-\phi)*\bar{V}}$

        $\rho\gets\sqrt{\frac{\rho_D}{\rho_V}}$ \tcp*{Modification factor}
        
        \If {$t \neq 1 \And \frac{\vert\bar{D}(t) - \Bar{\Bar{D}}\vert}{\bar{\bar{D}}} <= 0.5 $}{
        $\hat{\alpha} \gets \hat{\alpha} - \beta_2 * \bar{V}$ \tcp*{Reduce step size}
        
        $\hat{\alpha}\gets \max(\hat{\alpha},\max(\frac{\epsilon^{\mathrm{min}}_{\mathrm{t}}}{n},\frac{\bar{V}}{2}))$
        }
        \If {$t \neq 1 \And \rho > 1.5 $}{
        
        $\hat{\alpha}\gets \hat{\alpha}*{\rho}$\tcp*{Enlarge step size}
        }

    }
   \end{algorithm}
   A mobility adaptive gradient descent algorithm can be designed, as shown in Alg~\ref{alg:MAGD}. Step size is initialized in lines 5-6 with the given thresholds ($\epsilon^{\mathrm{max}}_{\mathrm{t_0}}$ and  $\epsilon^{\mathrm{min}}_{\mathrm{t_0}}$) at the beginning. As the localization error is initially large, a larger $\hat{\alpha}$ is used for stability. Lines 7-18 perform a gradient descent-based estimation using the previous position estimate (to guarantee a good convergence within $K$) and information from anchor \gls{uav}s. Momentum $m$ stabilizes gradient descent, while discount factor $\beta_1$ reduces step size within each estimation. $\Bar{D}_i$ represents the average distance difference between $\hat{d}_n$ and $d^{\circ}_n$, serving as an indicator of over-descending. Convergence threshold $\theta_t$ and max iteration threshold $K$ are used to terminate iterations. Lines 21-23 estimate current speed $V(t)$, average speed $\bar{V}$, and average distance difference $\bar{\bar{D}}$. The current distance difference $\bar{D}(t)$ can deviate due to occasional mis-localization, but a consistently enlarging $\bar{D}(t)$ suggests a small $\hat{\alpha}$, which results in a gradual loss of position accuracy. When $\hat{\alpha}$ is small, $\mathrm{p}(t)$ remains close to the previous estimate, resulting in a small estimated $V(t)$. 
   In lines 27-29, a smoothing window of length $\phi$ is applied to the estimates to mitigate fluctuations, and a modification factor $\rho$ is determined conjunctively by $V(t)$ and $\bar{D(t)}$ to modify $\bar{\alpha}$. Lines 24-28 adapt the learning rate based on stability and speed. When the position estimation is stable (indicated by $\Bar{D}(t)$ shows no significant deviation), $\hat{\alpha}$ is reduced by $\beta_2$ and $\bar{V}$ to ensure good accuracy. The minimum threshold of $\hat{\alpha}$ is determined by $\epsilon^{\mathrm{min}}_{\mathrm{t_0}}$ and $\bar{V}$. If the estimation is not stable, $\hat{\alpha}$ is increased by $\rho$.

   \subsection{Accuracy estimation of mutual localization system}\label{subsec:evaluofmu}
   We access the robustness and accuracy of \gls{magd} through simulations with varying numbers of anchor \gls{uav}s and fixed step sizes, and then compare the simulation results based on \gls{magd}. The movement speed of $u_k$ is periodically changed to simulate real mobility. Anchor \gls{uav}s $u_n$ are randomly initialized within a cubic area around target \gls{uav} $u_k$. The system configuration is summarized in Tab~\ref{tab:setup1}.

\begin{figure}[!htbp]
		\centering
		\includegraphics[width=0.90\linewidth]{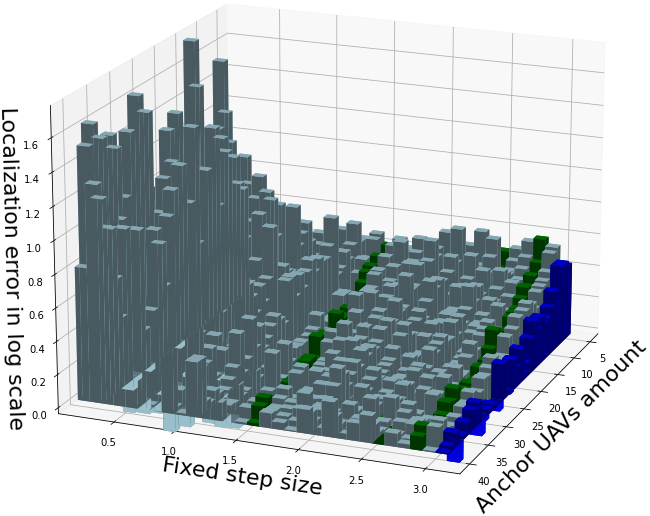}
		\caption{Localization error of fixed step sizes and \gls{magd} (3 best step sizes are marked in green and \gls{magd} in darkblue)}
		\label{fig:MAGDaccuracy}
	\end{figure}

   \begin{table}[!htbp]
		\centering
        \scriptsize
		\caption{Simulation Setup 1}
		\label{tab:setup1}
		\begin{tabular}{>{}m{0.2cm} | m{1.6cm} l m{3.7cm}}
			\toprule[2px]
			&\textbf{Parameter}&\textbf{Value}&\textbf{Remark}\\
			\midrule[1px]
			
            
			  &$n_a$&$5\sim40$& Anchor \gls{uav}s\\

			
			&$\sigma^2_{\mathrm{p},n}$&$\sim\mathcal{U}(0.1,3.0)$& Position error power / \si{\meter^2}\\
			&$V$&$\sim\mathcal{U}(0.6,3.4)$& Travel speed \si{\meter/s}\\
            & $T$ & 50 s &  Simulation time\\ 
            \midrule[1px]
            \multirow{-7}{*}{\rotatebox{90}{\textbf{System}}}
            &$[\epsilon^{\mathrm{max}}_{\mathrm{t_0}},\epsilon^{\mathrm{min}}_{\mathrm{t_0}}]$&[50,5]& Step size thresholds\\ 
            &$[\beta_1,\beta_2]$&$[0.5,0.05]$& Discount factors\\
			 
            &$m$&$1\times 10^{-5}$& Momentum\\
            &$\theta$&$1\times 10^{-8}$& Convergence threshold\\
			\multirow{-5}{*}{\rotatebox{90}{\textbf{MGAD}}}
            & $K$ & 30 &  Maximum iteration\\
            \bottomrule[2px]
		\end{tabular}
	\end{table}
     The simulation results in Fig~\ref{fig:MAGDaccuracy} depict the average error of 50 estimates. Among the three best fixed step sizes, $\alpha = (1.5,2.3,2.7)$ yield  average localization errors across all $n_a$ of $(1.63\metre,1.67\metre,1.66\metre)$ respectively. In comparison, the average localization error archived by \gls{magd} in this regard is 1.47\metre, indicating that \gls{magd} outperforms conventional approaches with fixed $\alpha^0$. In this specific setup, step sizes ranging from 1.4 to 3.0 show robust performance, although such a range may be challenging to find in practice. By adjusting $\alpha^0$ to different scenarios, our proposed method efficiently provides robust position estimation with good accuracy.   
    
    \section{Attack Paradigm and mutual attack detection system}\label{attack}
    \subsection{Attack Paradigm}{\label{attackpara}}
    Potential attacks on localization techniques can be categorized into several aspects.
  
    \textbf{Jamming mode}: the attacker jams the beacon signal of $u_n$ to introduce large distance estimation error and induce a wrongly received $\mathrm{p}(n)$ and $\sigma_{\mathrm{p},n}$. The received signal can be represented as $[\mathrm{p}^{\circ}_n + \mathrm{\tilde{p}}_n, ({d}^{\circ}_{k,n}+\tilde{d}_{k,n})^{+}, \sigma_{\mathrm{p},n} + \tilde{\sigma}_{\mathrm{J}}]$. Take the simplification, $\mathrm{\tilde{p}}_n\sim\mathcal{N}^3\left(0,\tilde{\sigma}_{\mathrm{J}}^2/3\right)$, $\tilde{d}_{k,n}\sim\mathcal{U}\left(0,\tilde{\sigma}_{\mathrm{J}}^2\right)$, where $\tilde{\sigma}^2_{\mathrm{J}}$ is the jamming index.
    \textbf{Bias mode}: the attacker hijacks some of the \gls{uav}s to erroneously report its position with a certain position bias to mislead others. In this scenario, the received signal can be modeled as $[\mathrm{p}^{\circ}_n + \tilde{\mathrm{B}},{d}^{\circ}_{k,n},\sigma_{\mathrm{p},n}]$, where $\tilde{\mathrm{B}}$ denotes position bias.
    \textbf{Manipulation mode}: the attacker hijacks some of the \gls{uav}s to report its position with an extra error, simultaneously modify its $\sigma_{\mathrm{p},n}$ to be extremely small for the intention of manipulating $w_n^e$. The received signal can be concluded as $[\mathrm{p}^{\circ}_n+\mathrm{\tilde{p}}_n, {d}^{\circ}_{k,n}, 1/\tilde{\sigma}_{\mathrm{M}}]$, where $\mathrm{\tilde{p}}_n\sim\mathcal{U}\left(0,\tilde{\sigma}^2_{\mathrm{M}}/3\right)$, $\tilde{\sigma}^2_{\mathrm{M}}$ is the manipulation index.
    
    In a constantly moving scenario, attack strategies can be categorized as follows. \textbf{Global random attack}: All malicious \gls{uav}s are uniformly distributed and randomly attack all nearby \gls{uav}s. This strategy aims to degrade position estimation globally and penetrate existing attack detection systems, as suggested in \cite{han2023trustawareness}. \textbf{Global coordinated attack}: Similar to the global random attack, the malicious \gls{uav}s coordinate their attacks within a certain time frame. \textbf{Stalking strategy}: All malicious \gls{uav}s follow a victim \gls{uav} and constantly attack it. This strategy does not impact estimation accuracy globally but targets the victim specifically.

    \subsection{Anomaly detection and trust propagation mechanism}
    Considering the above-mentioned attack schemes, a robust attack detection algorithm should be degrading the trustworthiness of suspicious \gls{uav}s. To achieve this, a reputation weight $r_n$ can be applied in \gls{magd}, more specifically in Alg~\ref{alg:MAGD} line 13 and 15,
    \begin{equation}
     G^i \gets \sum\limits_{u_n\in\mathcal{U}}\frac{(\hat{\mathrm{p}}-\mathrm{p}_n)*w_n^e*r_n }{d_n}*(d_n - d^{\circ}_n)
    \end{equation}
    \begin{equation}
    \bar{D}_i \gets \frac{1}{n} *\sum\limits_{u_n\in\mathcal{U}}(\hat{d}_n - d^{\circ}_n+\mu_{c,n})*w_n^e*r_n
    \end{equation}

    Our proposed method, illustrated in Alg~\ref{alg:TDAD}, estimates reputation weight using the cumulative distribution function of $\mathcal{N}(\mu_{\mathrm{cd},n},\sigma_{\mathrm{cd},n})$. In lines 7-8, $u_k$ calculates the estimated distance error $\hat{\mathrm{\mathcal{E}}}_n$ based on information from $u_n$ and the converted error distribution.
    Lines 12-16 address potential manipulation attacks by limiting $\sigma_{\mathrm{c},n}$ to a predefined minimum position error power $\sigma_{\mathrm{p},\min}$. Then compare the cumulative density $\xi_n$ with a preset probability threshold $\epsilon^t$ to detect the attack behavior of $u_n$. Depending on the detection results, the reputation weight update is performed with a punishment or reward $[\lambda_r,\lambda_p]$. The updated $r_n(t)$ considers its previous value $r_n(t-1)$, a forget factor $\gamma$, and the corresponding reward or penalty. Then $r_n(t)$ is thresholded within the range of [0,1].
    This approach aims at countering coordinated attacks, allowing $r_n(t)$ to recover gradually when attacks from $u_n$ become less frequent, without compromising mutual localization accuracy.
    
    \begin{algorithm}[!htbp]
    \caption{Time-evolving anomaly detection}
    \label{alg:TDAD}
    \scriptsize
    \DontPrintSemicolon
    Input: Reward $\lambda_r$ and penalty $\lambda_p$, 
			 forget factor $\gamma$ and confidence threshold $\epsilon^t$\\
    Output: $r_n(t)$
    
    Initialize: $r_n(t=1) \gets 1$
    
    
    \SetKwProg{Fn}{Function}{ is}{end}
    \Fn{\textsc{\gls{tad}}$(t = 1:T)$} {
        get $\hat{\mathrm{p}}(t)$ from MAGD$(t)$ meanwhile apply $r_n(t-1)$
        
        {    \For {$n = 1:N$}{get $\sigma_{\mathrm{p},n}, d^{\circ}_n, \mathrm{p}^{\circ}_n$ from $u_n$
        
                   get $\mu_{\mathrm{c},n},\sigma_{\mathrm{c},n}$
                   
                   
                   $\hat{d}_n \gets \Vert \hat{\mathrm{p}} - \mathrm{p}_n\Vert$ 
                   
                   $\hat{\mathrm{\mathcal{E}}}_n\gets \vert \hat{d}_n - d^{\circ}n + \mu_{\mathrm{cd}}\vert$ \tcp*{Distance error of $u_n$}
                   
                   calculate CDF $\xi_n$, 
                   
                    $\sigma_{\mathrm{c},n} \gets \max(\sigma_{\mathrm{c},n},\sigma_{\mathrm{p},min})$\\
                   $\xi_n \gets P_{\hat{\mathrm{\mathcal{E}}}_n} \left[\hat{\mathrm{\mathcal{E}}}_n~\vert~\mu_{\mathrm{c},n},(\sigma_{\mathrm{c},n})^2\right] $ 
                   
                   \uIf{$\xi_n > \epsilon^t$}{
                
                $\hat{r}_n \gets \lambda_r$ \tcp*{Assign reward}
               }\Else{
               
               $ \hat{r}_n \gets \lambda_p $ \tcp*{Assign penalty} }

             update $ r_n(t) $

             $ r_n(t) \gets \gamma*[r_n(t-1) + 1] - 1 + \hat{r}_n $

             $ r_n(t) \gets \min(1,\max(0,r_n(t)) $
             }      
    }}
    \end{algorithm}
    To address the vulnerability of target \gls{uav}s to spatial "ambushing" or stalking attacks, we incorporate a global reputation system. This system allows \gls{uav}s to share their local reputations with each other. Nevertheless, malicious reputation information can be shared as well, therefore a reputation propagation scheme is implemented. This ensures that reputation weights are carefully propagated, maintaining the reliability and accuracy of reputation information within the system. The propagated reputation weights can be described as follows:
    \begin{equation}
      \tilde{r}_{k,n}  = \frac{\sum\limits_{m\notin{k,n}}{r}_{k,m}*{r}_{m,n}}{\sum\limits_{m\notin{k,n}}{r}_{k,m}} , \tilde{\tilde{r}}_{k,n} = \frac{\mathrm{F_p}(\tilde{r}_{k,n})+{r}_{k,n}}{2}.
    \end{equation}
    While $u_k$ is utilizing the mutual localization system, $u_1...u_n$ are the accessible anchor \gls{uav}s, and ${r}_{k,n}$ is the local reputation. $u_m$ has uploaded its local reputation to the cloud, enabling reputation ${r}_{m,n}$ from $u_m$ to $u_n$ to be accessed. Meanwhile, $u_k$ has a local reputation ${r}_{k,m}$ to $u_m$. Based on the local reputation ${r}_{k,m}$, the uploaded reputation will be discriminated against accordingly. A propagated reputation $\tilde{r}_{k,n}$ is strongly leaning to \gls{uav} which has a good reputation to $u_k$. $\mathrm{F_p}$ is the propagation function designed to discriminate the already notorious $u_n$(a convex function is applied). Subsequently, proceed with the mean of local reputation and propagated reputation to \gls{magd}.   


    \section{Simulation results}\label{SIMUTDAD}
    \subsection{Evaluation under different attack mode and strategy}\label{subsec:globa}
       \begin{table}[!htbp]
		\centering
        \scriptsize
		\caption{Simulation Setup 2}
		\label{tab:setup2}
		\begin{tabular}{>{}m{0.2cm} | m{1.6cm} l m{3.7cm}}
			\toprule[2px]
			&\textbf{Parameter}&\textbf{Value}&\textbf{Remark}\\
			\midrule[1px]
			
                &Map Size&$[300,300,10]$& Define map size\\
                
			  &$n$&$100$& number of anchor \gls{uav}s\\
			
			&$\sigma^2_{\mathrm{p},n}$&$\sim\mathcal{U}(0.1,3.0)$& Position error power / \si{\meter^2}\\
			&$V$&$\sim\mathcal{U}(0.3,1.7)$& Travel speed \si{\meter/s}\\
            & $T$ & 100 s &  Simulation time\\
			\multirow{-5}{*}{\rotatebox{90}{\textbf{MGAD}}} 
            &$m$&$1\times 10^{-5}$& Momentum\\
            &$\theta$&$1\times 10^{-8}$& Convergence threshold\\
			
            & $K$ & 30 &  Maximum iteration\\
            
			\midrule[1px]

			&$[\lambda_r,\lambda_p]$ & $[0.3,-0.7]$& Reward and penalty\\
			&$\gamma$ & $0.5$& Forget factor\\

			&$\epsilon^t$ & $0.95$ & Confidence threshold\\
                \midrule[1px]
                \multirow{-6}{*}{\rotatebox{90}{\textbf{TAD}}}&$n_m$ & $30$& Malicious \gls{uav}s\\
                &$r_a$ & $0.5$&Attack rate if random attack\\
                &$T_a$ & $50$&Attack time frame if coordinated attack\\
                &$\tilde{\mathrm{B}}_n$& $[200,200,5]$&Position bias if bias attack\\
			&$\tilde{\sigma}^2_{\mathrm{M},n}$ & $200$& Manipulation index if manipulation attack attack\\
                \multirow{-7.5}{*}{\rotatebox{90}{\textbf{Attacker}}}&$\tilde{\sigma}^2_{\mathrm{J},n}$ & $5$& Jamming index if jamming attack 
                attack\\
            \bottomrule[2px]
		\end{tabular}
	\end{table}
    
\begin{figure}[!htbp]
		\centering
        \subfigure[Coordinated attack scheme]{
			\includegraphics[width=0.99\linewidth]{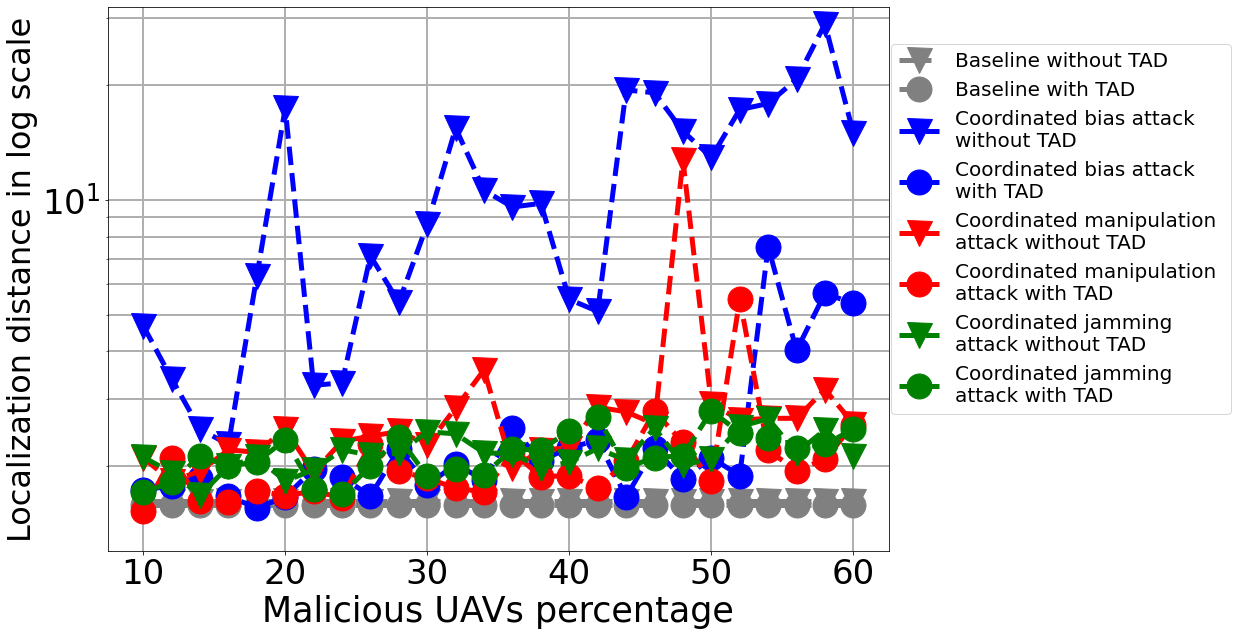}}
        \subfigure[Random attack scheme]{
		      \includegraphics[width=0.99\linewidth]{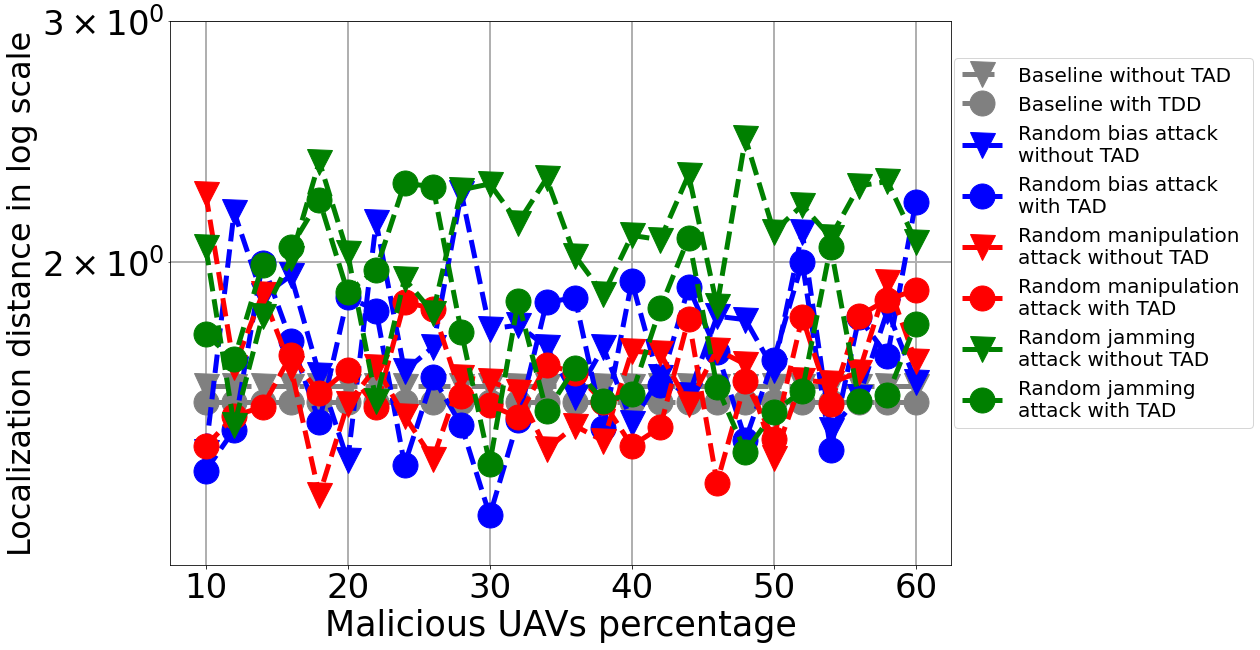}}
		\caption{Localization error of different over malicious \gls{uav}s percentage}
		\label{fig:localERR}
	\end{figure}

    \begin{table}[ht]
     \centering
     \caption{Estimate error ($\metre$) under different attack scheme and malicious \gls{uav}s percentage}
     \label{tab:accuracy1}
     \scriptsize
     \setlength{\tabcolsep}{5pt}
     \begin{tabular}{lccccc}
     \toprule
     \diagbox[width=10em,height=2.5em]{Attack schemes}{Percentage}& 10-20\%  & 20-30\% & 30-40\% & 40-50\% & 50-60\%\\
     \midrule
      No attack  &  1.62 & -- & -- & -- & --\\
      No attack with \gls{tad}  &  1.58 & -- & -- & -- & --\\
      Coord. Bias & 3.89  & 7.29 & 10.80& 12.84 & 19.54 \\
      Coord. Bias with \gls{tad} & 1.72  & 1.88 & 2.05& 2.06 & 4.25 \\
      Random Bias & 1.82  & 1.84 & 1.73 & 1.68& 1.74 \\
      Random Bias with \gls{tad} & 1.64  & 1.66 & 1.62& 1.74 & 1.74 \\
      Coord. Mani. & 2.05  & 2.32 & 2.58 & 4.63 & 2.82\\
      Coord. Mani. with \gls{tad} & 1.72  & 1.86 & 1.88& 2.16 & 2.71 \\     
      Random Mani. & 1.77  & 1.57 & 1.54 & 1.68 & 1.66\\
      Random Mani. with \gls{tad} & 1.58  & 1.71 & 1.61& 1.56 & 1.71 \\
      Coord. Jam. & 1.97  & 2.06 & 2.24 & 2.18 & 2.39\\
      Coord. Jam. with \gls{tad} & 1.94  & 2.03 & 2.03& 2.28 & 2.43 \\
      Random Jam. & 1.95  & 1.93 & 2.12 & 2.16 & 2.19\\
      Random Jam. with \gls{tad} & 1.95  & 2.04 & 1.62 & 1.72 & 1.68\\
      \bottomrule
      \end{tabular}
    \end{table}
    To validate the robustness and effectiveness of \gls{tad}, we carried out simulations of our mutual localization system under different attack modes and strategies, while 10 targets $u_k$ navigates through attackers. We assume the attacker has limited resources and can only jam/hijack some part of the \gls{uav}s. The attacks can be organized as random or coordinated. The \gls{magd} configuration follows Tab.\ref{tab:setup1}, while system, \gls{tad} and attacker parameters are listed in Tab.\ref{tab:setup2}. The attack rate and attack time frame are set to 0.5 and 50 to keep an unvarying attack power. All anchor \gls{uav}s are distributed within the map and navigate to random destinations, which leads to an average of 15 anchor \gls{uav}s within a mutual localization range of $50\metre$. Target \gls{uav}s can be anchor \gls{uav}s to each other, $\sigma^2_{\mathrm{p},k}$ is set to be $\max(\sigma^2_{\mathrm{p},n})$. The simulation results are shown in Fig.~\ref{fig:localERR} and  Tab.~\ref{tab:accuracy1}.
 
    Contrary to findings in static sensor networks, our results indicate that coordinated attacks are more effective than uncoordinated ones, contradicting the suggestion that both schemes are equally effective \cite{uncoordinated}\cite{Ragmalicous}. The inconsistency in error injection during each \gls{magd}'s gradient descent process leads to this difference. \gls{magd} can still provide accurate estimations under less dense attacks due to varying attack density, preventing a cascading effect of mislocalization caused by injected errors. Random attacks lead to temporary large localization errors, which can be quickly resolved within a few timesteps. In contrast, coordinated attacks cause sustained mis-localization (detailed trace error analysis emitted in this paper due to the length limit). Moreover, the results demonstrate that attack mode bias exhibits higher effectiveness in causing large localization errors. This is attributed to the fact that the injected errors in this mode do not offset each other, leading to a cumulative impact on the localization accuracy. Additionally, the coordinated jamming mode poses a formidable challenge to \gls{tad} as it enlarges $\sigma_{c,n}$ of malicious \gls{uav}s, which makes it challenging for \gls{tad} to identify attacks. Nonetheless, our proposed weighted localization approach effectively mitigates such attacks by favoring anchor \gls{uav}s with smaller $\sigma_{c,n}$, thereby bolstering the overall resilience and robustness of the localization system. 

\begin{figure}[!htbp]
		\centering
		\includegraphics[width=0.90\linewidth]{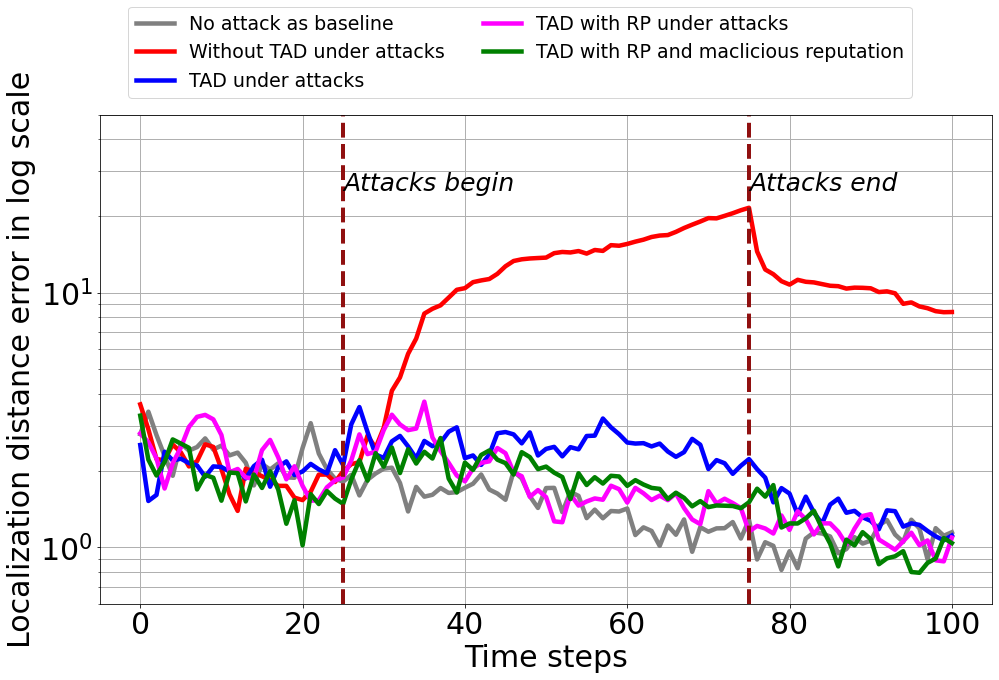}
		\caption{Localization error at different time step (for the baseline, UAVs are set to follow the target as well)}
		\label{fig:ERRinTP}
	\end{figure}

    To further assess \gls{tad} and \gls{rp}, we conducted simulations with malicious \gls{uav}s employing a coordinated stalking strategy to attack the victim target $u_k$. The parameters for \gls{magd} and \gls{tad} follow those listed in Table~\ref{tab:setup2}, while the attack mode is set to bias mode. The percentage of malicious \gls{uav}s is set at $30\%$, which includes 3 target \gls{uav}s capable of attacking other target \gls{uav}s while sharing malicious reputation. We assume the attacker knows the actual position of the victim target with ambiguity (the estimated position $\hat{\mathrm{p}}_k$, shared by $u_k$, should not be used as it can be misleading when attacks become effective). The results in Fig.~\ref{fig:ERRinTP} show average errors of 20 simulations at different time steps.  
 
   Our simulation results demonstrate the effectiveness of \gls{tad} against stalking attack strategies with the given attack density, significantly reducing localization errors compared to the absence of \gls{tad}. Moreover, the introduction of RP further enhances localization performance, leading to a faster convergence of localization errors compared to \gls{tad} alone. Notably, RP also exhibits resilience to malicious reputation information.

    \section{Conclusion}\label{conclu}
    In this paper, we evaluated localization techniques and presented a novel localization scheme (\gls{magd}) that adapts to the mobility and changing availability of anchor \gls{uav}s. Additionally, we introduced defense schemes (\gls{tad} and RP) to counter potential attacks. Our numerical simulations demonstrated the effectiveness of these methodologies in dynamic scenarios.

    However, it is essential to acknowledge that this study did not extensively address potential attacks against \gls{rp}. A sophisticated attacker might manipulate a subset of compromised \gls{uav}s to launch attacks on target \gls{uav}s while others share a malicious reputation. This poses a challenge for our \gls{tad} and RP to effectively detect and mitigate such attacks. The potential threat calls for a novel approach to identifying attack patterns. 
    \section*{Acknowledgment}
	This work is supported in part by the German Federal Min-
    istry of Education and Research within the project Open6GHub (16KISK003K/16KISK004), in part by the European Commission within the Horizon Europe project Hexa-X-II (101095759). B. Han (bin.han@rptu.de) is the corresponding author.

\bibliographystyle{IEEEtran}
\bibliography{references}

\end{document}

%% file: glossary.tex
\makeglossaries
\newacronym{uav}{UAV}{unmanned aerial vehicle}
\newacronym{magd}{MAGD}{Mobility Adaptive Gradient Descent}
\newacronym{tad}{TAD}{Time-evolving Anomaly Detectio}
\newacronym{gps}{GPS}{Global Positioning System}
\newacronym{tof}{ToF}{Time of Flight}
\newacronym{rssi}{RSSI}{Received Signal Strength Indicator}
\newacronym{mle}{MLE}{maximum likelihood estimator}
\newacronym{rp}{RP}{reputation propagation}
\newacronym{ls}{LS}{least square}
\newacronym{ln-1}{LN-1}{$l$1 norm}
\newacronym{gd}{GD}{gradient descend}
\newacronym{admm}{ADMM}{Alternating
Direction Method of Multiplier}